\documentclass[prl,aps,twocolumn,showpacs,floatfix,superscriptaddress]{revtex4}

\usepackage{graphicx,times}
\newcommand{\MEMO}[1]{}
\newcommand{\KEEP}[1]{{#1}}
\newcommand{\OMIT}[1]{}

\newcommand{\SAVE}[1]{}
\def\secc#1{\smallskip\underline{\it #1} ---}

\begin{document}
\title{Empirical oscillating potentials for alloys from ab-initio fits}

\author{Marek Mihalkovi\v{c}}
\affiliation{Laboratory of Atomic and Solid State Physics, Cornell University,
Ithaca, NY, 14853-2501}
\affiliation{
Permanent address,
Institute of Physics, Slovak Academy of Sciences, 84228 Bratislava, Slovakia.}
\author{C. L. Henley}
\affiliation{Laboratory of Atomic and Solid State Physics, Cornell University,
Ithaca, NY, 14853-2501}
\author{M. Widom}
\affiliation{Dept. of Physics, Carnegie-Mellon University,
Pittsburgh, PA , 15213}
\author{P. Ganesh}
\affiliation{Dept. of Physics, Carnegie-Mellon University,
Pittsburgh, PA , 15213}
\affiliation{Current address: Carnegie Institution of Washington, GL,
5251 Broad Branch Road, NW, Washington, DC, 20015}

\begin{abstract}
{By fitting to a database of ab-initio forces and energies, 
we can extract pair potentials for alloys, with  a simple six-parameter
analytic form including Friedel oscillations, which give
a remarkably faithful account of many complex intermetallic
compounds.  As examples we show results for (crystal or
quasicrystal) structure prediction
and phonon spectrum for three systems: Fe--B, Al--Mg--Zn, 
and Al--Cu--Fe.}
\end{abstract}


\maketitle

Several kinds of problem in materials modeling can be 
addressed only by classical interatomic potentials.
Consider complex alloy crystal structures:
\OMIT{(e.g. Al-transition metal)}
it is impossible to use diffraction-data
refined structures straightforwardly, for they almost always 
contain sites with mixed or fractional occupancies.
These must be resolved properly to evaluate physical properties, 
which can be unfeasible with current fast ab-initio codes 
using density functional theory, such as VASP~\cite{vasp}.
Only moderate resources are needed for {\it single} evaluations 
of the total energy, but repeated evaluations
are prohibitive for unit cells having up to $10^3$ atoms per cell;
even if just 1--5\% of the sites are uncertain,
one must examine a vast number of variant structures 
to assign the occupancies optimally.  
In any case, classical potentials 
are required when the structure has no tractable unit cell
(quasicrystals or amorphous metals); they also facilitate
dynamic or thermodynamic simulations (phonon spectra and
phase transformations in complex alloys).

\SAVE{Not mentioned: Material design/modelling, bridging 
microscopic-to-macroscopic scales. Nanocrystals.
As for dislocations, maybe not valid.}

A common modern approach to modelling atomic interactions 
classically is the ``embedded-atom method'' (EAM)~\cite{EAM}
as well as ``modified EAM''~\cite{MEAM},
in which the full Hamiltonian contains the usual pair term $V_{ij}(R)$, 
but also an implicitly many-atom term $U(\rho)$, where $\rho$ 
is a sum of contributions from nearby atoms.
Accurate EAM potentials are straightforward to extract
for monatomic systems~\cite{EAM-element}
but demand patience and skill to obtain even for binary systems
~\cite{EAM-alloy,gaehler-brommer}; obviously, dimensionality of
parameter space becomes critical for multicomponent systems.

\SAVE{$\rho$ is given by the ``embedding functions'' $\rho_{i}(R_1, ..., R_n)$. 
EAM potentials are fitted to 
ab-initio data (forces, energies...) either as analytical 
forms,  or (the so-called ``glue'' potentials) 
are represented (using splines) as arbitrary functions sampled
at intervals of radius.''}

In this paper we report on an alternative approach~\cite{mm-review}
fitting only pair interactions but incorporating Friedel oscillations,
optimized (typically) for a particular composition range.
The oscillating analytical form is natural only for simple 
metals (Al, Mg,...)  yet (see below)
it works surprisingly well even when 
angular or many-body interactions are important.
\OMIT{such as glass-forming B-Fe alloys.~\cite{ganesh-BFe}.}
We will first describe the form of the potential and our
methods for fitting it, then 
demonstrate its capabilities 
through case studies in three alloy systems: Fe-B, 
Al-Mg-Zn, and Al-Cu-Fe; 
finally, we will  summarize other systems where this method has been applied
and discuss its limitations.


\secc{Method: potential, database, and fitting}
Our ``empirical oscillating pair potentials'' (EOPP)
have the form
   \begin{equation}
   \label{eq:oscil6}
    V(r) = \frac{C_1}{r^{\eta_1}} + \frac{C_2} {r^{\eta_2}} \cos(k_* r + \phi_*)
   \end{equation}
\SAVE{Relation to Marek's internal form (with $\{ a_i \}$):
$C_1=a_0^{a_1}$; $\eta_1=a_1$; 
$C_2=a_2$; $\eta_2=a_5$;
$k_*=a_3$; $\phi_*=a_4$.}
All six parameters, including \OMIT{the spatial frequency} $k_*$,
are taken as independent in the fit for each pair of elements. 
Eq.~(\ref{eq:oscil6}) was inspired by effective potentials
(e.g. \cite{hafner} and \cite{widom-GPT}) 
used in previous work
on structurally complex metals, e.g.
quasicrystals~\cite{Al-TM,krajci-AlZnMg,mm-review}.
In such systems, energy differences between competing structures
are often controlled by second- and third-neighbor wells due to
Friedel oscillations, 
\KEEP{which are a consequence (mathematically)
of Fourier transforming the Fermi surface, or (physically) are
equivalent to the Hume-Rothery stabilization by enhancing
the strength of structure factors that hybridize states
across the Fermi surface.}%
~\cite{pettifor}
In the framework of Ref.~\onlinecite{pettifor} (Sec. 6.6),
the short-range repulsion is captured by the first term of
(\ref{eq:oscil6}); the medium-range potential (first-neighbor well) 
as well as the long-range oscillatory tail are captured by
the second term, their relative weights being adjusted by
the $\eta_i$ parameters.
Note that empirically fitted potentials account for some of
of the many-body contributions within their pair terms, 
which works better practically than truncating a
systematic expansion (e.g. GPT~\cite{moriarty-GPT})
after the pair terms. 

\SAVE{Name 
Nomenclature: ``Empirical oscillating'' pair potential (EOPP).
including ``pair'' to distinguish from e.g. G\"ahler's 
fitted EAM potentials with Friedel oscillations; 
including ``empirical'' to distinguish e.g. from GPT.}

\SAVE{Regarding the accounting of greater-than-two atom terms:
the possibility of folding some portion of
the higher order terms into the pair potential was discussed
in J. A. Moriarty, Phys. Rev. B 42, 1609 (1990); 49, 12 431 (1994).}

\SAVE{
A difference between our parametrization and Pettifor's is that
our repulsion term is not exponential.  CLH notes that analytically
the long-range tail has a fixed power -- $\eta_2=3$ and $k_*=k_F$ --
but (as stated in above para.) the effect of fitting $\eta_2$ is to 
allow the first well to vary independently of the others.  
(Also, in our range of radii, the expected analytic potential
has subdominant terms in higher powers of $1/r$, so the fitted
$\eta_2$ should deviate in any case.)  
Or perhaps the nearest-neighbor well gets
created by competition of the first and second terms.  
Thus, the medium-range part is not missing, just hidden.
In some cases, e.g. Al-Al in many systems 
(see Tables in the deposited supplement), the effective potential 
is purely repulsive correponding to small $C_2$ or large $\eta_2$.
In other cases, the fitted hardcore is small
(small $C_1$ or $\alpha_1$),
the nearest-neighbor well being accounted by the oscillating term.
[The hardcore repulsion in Eq.~(\ref{eq:oscil6}) is trustworthy, of course, 
only down to $r$ explored by the MD simulations; but this suffices for the
closest pairs in the equilibrium state of complex structures 
(e.g. Al-Co in a quasicrystal).]}

\begin{table}
\begin{tabular}{|l|c|c|c|}
\hline
system        &  B-Fe  &  Al-Mg-Zn  & Al-Cu-Fe   \\
\hline
ref. & Fe,\ BFe$_2$ & Al,\ Mg,\ Zn &Al$_7$Cu$_2$Fe\\
        &  &  & Al$_2$Cu,\ Al$_2$Fe  \\
\hline
data- & BFe$_2$ [Al$_2$Cu]   & MgZn$_2$ [Laves]  &  Al$_6$Fe      \\
base        & BFe$_3 $ [CFe$_3$]    & $T$(Al$_{96}$Mg$_{64}$) & Al$_5$Fe$_2$ [Al$_{23}$CuFe$_4$] \\
        & BFe$_3 $ [Ni$_3$P]    & \quad [$T$(AlMgZn)]    &  Al$_2$Cu \\
        & B$_6$Fe$_{23}$ [C$_6$Cr$_{23}$] & $T$(Zn$_{96}$Mg$_{64}$) & Al$_7$Cu$_2$Fe\\
        & B$_3$Fe$_4$ [B$_3$Ni$_4$]  & \quad [$T$(AlMgZn)]    & Al$_{23}$CuFe$_4$  \\
        & B$_1$Fe$_1$ [BCr]     & Al$_{12}$Mg$_{17}$      &  Al$_{72}$Cu$_4$Fe$_{24}$ [Al$_3$Fe] \\
        & $a$-Fe$_{80}$B$_{20}$  & AlMg$_4$Zn$_{11}$       & ``1/1-expe''  \\
        &                       & $\beta'$-Al$_3$Mg$_2$   & ``1/1-6D''    \\
\hline
$T_{\rm MD}$ & 1500 & 450 (Al-Mg);  &  500 (``1/1-6D''); \\
                 &      &  1000 (Mg-Zn) & 1200 (``1/1-6D''); \\
                 &      &               & 1000 (Al$_7$Cu$_2$Fe) \\
\hline
\end{tabular}
\caption{Compounds used in each database used to fit potentials.
Energies are expressed as difference from the tie-line or tie-plane
defined by the simple reference structures listed; these structures
are also included in the database.
In brackets are the conventional names for crystal structure type.
The temperature (in K) used for molecular-dynamics samples
is $T_{\rm MD}$ (where compositions are given, those are the
only $T>0$ inputs)}
\MEMO{NEED(1):  CLH must check Marek's table of 
relaxed and which used at high $T$, and find a short way to mark here.
}
\label{tab:database}
\end{table}

\SAVE{Al-Mg-Zn source is Marek's fit ``set2/energy/001'' .}

The parameters in (\ref{eq:oscil6}) are fitted to an ab-initio dataset
(we always used the VASP code~\cite{vasp})
combining both relaxed $T=0$ structures and molecular dynamics 
(MD) simulations at high $T$ (usually the same structures,
below the melting point.)
Structures are selected for the database within, or bracketing,
the composition range of interest; when possible,
a mix of simple and complex structures is used. 
A key criterion in choosing
structures is to ensure an adequate number of contacts of
each kind (in particular, nearest-neighbor contacts between 
the least abundant species). 
Also, all structures in the database should have 
similar atom densities~\cite{fn-densityrange}.
In particular, our high-$T$ MD samples
were constrained to have the {\it same} density as at $T=0$,
rather than the physical zero-pressure values.
In MD simulations of the simpler structures, a
supercell is always used with dimensions comparable to the
potential cutoff radius, which (in this paper) is  always 12\AA
\SAVE{and abrupt}.

\SAVE{A database has four quadrants: energies/forces $\times$ 
relaxed/high T.  Although the $T=0$ ab-initio forces are
of course zero, they are just as important to include in
the simulation: the pair-potentials aren't necessarily zero
for those configurations!}

\SAVE{(MM): A warning: if our database has 
[structures with a] nonuniform electron density,
this  makes the fit less reliable.  For example, 
we might attempt to assess the vacancy formation
energies by varying site occupancies.
The potentials could fit this by adjustments of the
first-neighbor well depth or width;
but, very likely, this will be inconsistent
with the forces at first-neighbor distances;
so the overall fit would fail.
[Presumably, a truly consistent fit
of vacancy formation energies depends on multi-atom
interactions and needs something like EAM.]
Yet there are cases when the fit works reasonably:
e.g. the Al-V potential in Al-rich limit (using samples with 
a site empty or occupied by Al).
[MM also noted (12/07)
that when GPT potentials are not used under fixed-volume conditions
they were derived for, a small difference in oscillating part 
usually leads to large RMS values; I presume the same problem
can happen for EOPP.]}

\SAVE{
The fit uses both ab-initio forces and energy differences. 
We have implemented two different kinds of fits: 
There is another mode of fitting in which  energy
datapoints are always differences between different samples
at the same composition.  This is good when the purpose
is to accurately tune the forces (e.g. before doing a 
phonon calculation), but it isn't represented among the
projects reported here.}

\begin{figure}
\vskip 1.0cm   
\includegraphics[width=3.2in]{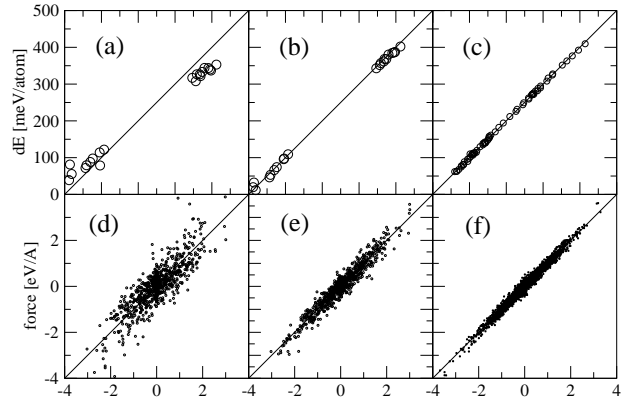}
\caption{Scatter plot of pair potential result (vertical
axis) versus ab-initio data (horizontal axis) for
energy differences $\Delta E_i$ (top panels)
and forces ${\bf F}_j$ (bottom panels)
The potentials are:  (a,d) Lennard-Jones fit to B-Fe,
(b,e) EOPP fit [Eq.~(\ref{eq:oscil6})] to B-Fe, and 
(c,f) EOPP fit to Al-Mg-Zn.}
\label{fig:scatter}
\end{figure}

We define each structure's energy as a difference
relative to a coexisting mixture (with the same 
total composition) of reference phases,
chosen to bracket all database compositions.
\SAVE{Depending what tie-plane you
choose, you obtain different potentials.  
MM says that, by choosing the tie-plane to bracket the
database (usually, pure elements), you reproduce the energy part
of the database better.
(What would implement a change of chemical potentials?  Does a 
change in tie-plane do that?)
MW: ``Rather than adding constants, its better to think about adding a
localized term that alters the near-neighbor energy without
significantly altering the near-neighbor force. For pure elements
this amounts to a chemical potential shift but for mixed
potentials it governs the enthalpy of mixing.
Of course the potentials [as in (\ref{eq:oscil6})]
need to [asymptote] to zero to assure convergence of the energy.''}
Every structure is used for both forces (from MD at high $T$) and
energy differences~\cite{FN-need-energies}
(high-$T$ MD, as well as relaxed at $T=0$).
For the high-$T$ portion, we took one snapshot of each structure
at the end of a short ab-initio MD run.
Typically $\sim 10^3$  force components entered the fit, 
along with $\sim 50$ energy differences (more for Al-Mg-Zn, 
fewer for Fe-B), and
\OMIT{(see e.g. Fig.~\ref{fig:scatter})}
the forces are $\sim 2$--4 eV/\AA~ while the energy differences
are $\sim 0.2$--0.4 eV/atom; the fit residuals are
$\sim 5$\% and $\sim 1$\% respectively.

Our least-squares fit minimizes (by the Levenberg-Marquardt algorithm)
     $\chi^2 \equiv \sum \Delta E_i^2/ {\sigma_E}^2 + 
     \sum |\Delta {\bf F}_j|^2/{\sigma_F}^2$, 
where $\{ \Delta E_i \}$ and $\{ \Delta {\bf F}_j \}$ are the
energy and force residuals; we found a weighting ratio
$\sigma_E/\sigma_F \sim 10^{-3}$\AA~ was optimal so
that neither energies nor forces dominate the fit.
\SAVE{CLH notes: presumably the appropriate ratio depends on the
no. of high-$T$ data points relative to no. of relaxed?
The reason we needed to incorporate (at least a little bit) energy data, 
in the first place, is that some parameter combinations
are ill-determined by forces.}
There is some risk of converging to a false minimum
(or not converging at all, from an unreasonable initial guess).
Thus, it is important to repeat the fit from several starting guesses.  
For this we used, e.g., potentials first fitted to pure elements
or binary systems, and also used a library of parameter sets
\OMIT{(such as in Table ~\ref{tab:6param})}
previously fitted for some
different alloy system.
The fitted parameters in (\ref{eq:oscil6}) for each of our 
examples are gathered in the table ~\ref{tab:6param};
similar potentials were plotted in Refs.~\cite{CMA-AlMg} (for Al-Mg) 
and \cite{boissieu-ScZn} (for Sc-Zn).

\SAVE{In Al--TM systems, GPT potentials were excellent initial guesses
for fitted potentials. 
[very often few fitting iterations dramatically improve the fit.]}

\secc{Example 1: B-Fe. }
Amorphous B-Fe is the simplest member of
a family of technologically important metallic glasses.
Our B-Fe database (Table ~\ref{tab:database}) included
several crystals plus an ``amorphous'' sample with 100 atoms in 
an approximately cubic box.
The ab-initio data were calculated with a spin polarization
so as to include magnetic contributions to the 
forces/energies.~\cite{ganesh-BFe}
For comparison, we also fitted the database to  Lennard-Jones 
(LJ) potentials which lack oscillations (starting 
this fit from the Kob-Anderson potentials~\cite{kob-and}).

\SAVE{(MW, 1/08)
For ab-initio specialists, "spin polarization" implies uniform
direction. The magnitude [here] is not uniform and it is not restricted
to Fe atoms.}

\SAVE{(Ganesh, 1/08):
Both the $T=0$ energy and forces ($\langle F\rangle = 0.0403$,
$|F_{max}|= 0.1111$) were included for a-FeB.
For MD of B$_{20}$Fe$_{80}$ at T=1500K, 11 energy samples 
and 2 force samples were included;
For MD of BFe3.oP16 at T=1000K where the atoms have undergone slight
displacements from the equilibrium positions in the crystal phase, 5 energy
samples and 2 force samples were included.
MM 1/7/08 : turned to SAVE
}

\SAVE{CLH has not incorporated the clarifications in
Ganesh's email of 1/3/08}

We employed a modified tempering scheme~\cite{lyman} with replica exchange.
That means two simulations were carried out
for 100-atom samples of Fe$_{80}$B$_{20}$ at T=1500K, one using 
the ab-initio energy as its Hamiltonian and the other using EOPP.
\SAVE{Time steps for the first-principles (FP) calculation
require more time than EO time steps.}
At intervals a Monte Carlo (MC) attempt is made to
swap the two samples.
These swaps accelerate the configuration sampling.
The acceptance probability (we had $\sim 70$\% for Fe-B) is 
a stringent test of how well the pair-potential
Hamiltonian mimics the ab-initio one.
\OMIT{The resulting ensemble can then be used
as the basis for tempering (temperature exchange) MD
to efficiently sample lower temperatures.}

\begin{figure}
\includegraphics[width=3.2in]{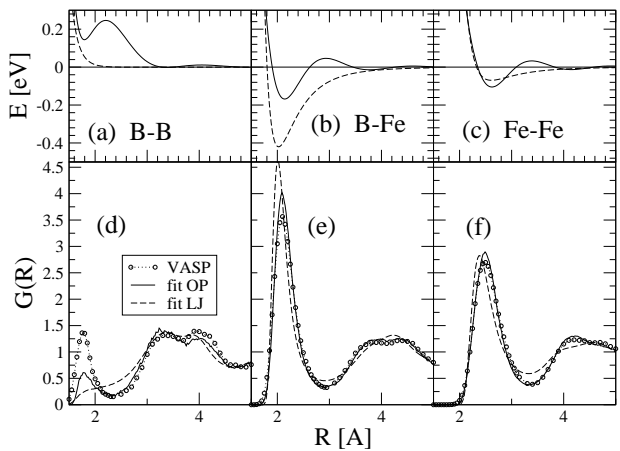}
\caption{a,b,c: B-Fe pair potentials (c,d,e): Corresponding
radial distribution functions at 1500K, using ab-initio energies
(dotted), empirical oscillating pair-potentials (solid), or
fitted Lennard-Jones potentials (dashed).
}
\label{fig:glass}
\end{figure}

The superiority of the EOPP form over LJ is evident from
the excellent diagnostic of scatter plots  in
Fig. \ref{fig:scatter}.
\OMIT{(compare the leftmost versus the middle panels)}
As another diagnostic, 
Fig.~\ref{fig:glass}, shows
the radial distribution
functions found in an MD run at $T=$1500K, 
using the ab-initio energies 
as well as both kinds of pair potential;
the EOPP form faithfully reproduced all features of the ab-initio data.
\SAVE{I have a note``Lower-left corner in 
Fig.~\ref{fig:scatter}(b) has an outlier: the 
EOPP energy of the B$_6$Fe$_{23}$ crystal is too high.}

\secc{Example 2: Al-Mg-Zn} 
In this example, EOP potentials resolve site
occupancies and the phase diagram of low-temperature modifications
of the well-known cubic ``Bergman'' compound T(AlMgZn)~\cite{bergman,mizutani}
(X-rays can hardly distinguish Al from Mg ions, as they differ
by only one electron.)
Our database (Table~\ref{tab:database}) spanned
the entire range of ternary compositions.
(Al--Mg--Zn was previously modeled using 
pseudo-binary pair potentials,~\cite{krajci-AlZnMg}.)
\OMIT{(in which Al/Zn are treated as one species.)}

\OMIT{
\begin{table}
\begin{tabular}{|c|cccccc|}
\hline
pair   &  $C_1$  \ & $\eta_1$ \ &  $C_2$ \ &  $\eta_2$ \ & $k_*$ \ &  $\varphi_*$ \ \\
\hline
Al-Al \ &\    2669 \ &  10.37 \ &  -4.93 \ &  4.41  \ &  3.88 \ &  -4.38 \ \\
Al-Mg \ &\    4296 \ &  10.61 \ &  -39.4 \ &  6.21  \ &  3.33 \ &  -4.02 \ \\
Mg-Mg \ &\   14.59 \ &  4.79 \ &  183.3 \ &  6.49  \ &  2.19 \ &  -4.53 \ \\
Al-Zn \ &\     949 \ &  9.36 \ &  -2.52 \ &  3.36  \ &  3.33 \ &  -2.16 \ \\
Mg-Zn \ &\     282 \ &  8.26 \ &  -31.7 \ &  5.58  \ &  2.82 \ &  -2.26 \ \\
Zn-Zn \ &\     706 \ &  9.33 \ &  -6.01 \ &  4.33  \ &  3.69 \ &  -3.48 \ \\
\hline
\end{tabular}
\caption{Parameters for AlMgZn ternary potentials [Eq.~(\ref{eq:oscil6})],
with $r$ in units of \AA~ and $V(r)$ in eV.}
\label{tab:6param}
\end{table}
}


\begin{table}
\begin{tabular}{|c|cccccc|}
pair & $C_1$ & $\eta_1$ & $C_2$ & $\eta_2$ & $k_*$ & $\varphi_*$ \\ 
\hline
B--B &  11.69 & 5.184 & 8.338 & 5.149 & 3.950 & 2.617  \\
B--Fe &  2372.51 & 17.232 & -6.043 & 4.441 & 3.994 & 3.637  \\
Fe--Fe &  1.443$\times 10^5$ & 18.556 & 4.939 & 3.934 & 4.074 & 4.651  \\
\hline
Al-Al  &    2669  &  10.37 &  -4.93  &  4.41   &  3.88  &  -4.38  \\
Al-Mg  &    4296  &  10.61 &  -39.4  &  6.21   &  3.33  &  -4.02  \\
Mg-Mg  &   14.59  &  4.79  &  183.3  &  6.49   &  2.19  &  -4.53  \\
Al-Zn  &     949  &  9.36  &  -2.52  &  3.36   &  3.33  &  -2.16  \\
Mg-Zn  &     282  &  8.26  &  -31.7  &  5.58   &  2.82  &  -2.26  \\
Zn-Zn  &     706  &  9.33  &  -6.01  &  4.33   &  3.69  &  -3.48  \\
\hline
Al--Al &  4.431$\times 10^5$ & 15.990 & -0.577 & 3.410 & 4.680 & 5.372  \\
Al--Fe &  1.842$\times 10^5$ & 18.713 & 7.884 & 3.609 & 3.079 & 1.730  \\
Al--Cu &  3096.1 & 12.079 & -1.457 & 3.261 & 3.687 & 2.978  \\
Fe--Fe &  125.26 & 8.081 & 1.115 & 2.504 & 4.477 & 1.708  \\
Fe--Cu &  23.13 & 5.856 & 0.323 & 1.520 & 3.086 & 1.947  \\
Cu--Cu &  11304 & 12.631 & -2.124 & 3.301 & 2.830 & 0.159  \\
\hline
\end{tabular}
\caption{Parameters for B--Fe, Al--Mg--Zn and Al--Cu--Fe potentials [Eq.~(\ref{eq:oscil6})],
with $r$ in units of \AA~ and $V(r)$ in eV.}
\label{tab:6param}
\end{table}


\SAVE{
Our Al-Mg fitted EOP was used in Ref.~\cite{CMA-AlMg}
to resolve ambiguities in the $T=0$ optimal structure.}

\SAVE{(see CLH emails 8/9/07)
CLH worries: since the Al-Mg-Zn crystals we are solving for, 
also got included in the database, it may make the errors look
spuriously good.
MW replied: I don't see anything intrinsically wrong
with including the data in the fit, as long as the procedure is
accurately described in the paper.  A really
impressive test would be to predict values
that were not included in a fit.}

The EOP potentials were then used in a lattice-gas type MC 
annealing (using sites from diffraction~\cite{bergman,mizutani})
for each of 91 compositions,
with 52--64 Mg atoms and 24--72 Al atoms per cell
(thus spanning the full Al-Zn range, and $\pm$5\% in Mg concentration).
For 21 selected candidate structures, we 
subsequently did a full  ab-initio calculation. 
Four of these structures (Table~\ref{tab:almgzn})
were stable at $T=0$.
\SAVE{i.e. they occur on the convex hull of the 
scatter plot of energy vs composition.}
The first of these is the standard structure~\cite{bergman}
(but with empty-centered icosahedral clusters); 
the second one differs by converting
the Mg atoms with coordination 14 (the smallest for Mg)
to Al atoms.
We applied the same method
to the recently solved $\beta'$(AlMg) structure~\cite{CMA-AlMg},
a low-$T$ variant of Samson's very large-cell 
$\beta$(AlMg)~\cite{samson-beta-AlMg};
at $T=0$ $\beta'$(AlMg) 
is essentially a line compound.
\OMIT{occupying just $\sim 1$\% of the composition interval.}
\SAVE{Binary Al-Mg potentials could be 
fitted accurately throughout the whole compositional range!''}

\SAVE{
MM notes: these potentials account well for the
unpublished phonon dispersion relation 
for Al-Mg-Zn, 
recently measured (fall '06) by M. de Boissieu and
M. Mihalkovi\v{c}.
More comparison could be made with Hafner's or Moriarty's potentials, 
though this will be complicated by conditions (concentration [?] bracketing,
relative weighting forces/energy).  
The case of pure Mg shows the best agreement:
Moriarty's potential isw virtually indistinguishable from the fitted
version.}

\begin{table}
\begin{tabular}{|c|ccc|c|cc|}
\hline
composition      &   Al(2) &  Mg(3) &  Zn(1) & sp. gr.& 
     $\Delta E_{\rm VASP}$  & $\Delta E_{\rm pair}$ \\
\hline
Al$_{24}$Mg$_{64}$Zn$_{72}$ & 1  & 0   & 0    & $Im\bar{3}$ &-122.1&-117.3 \\  
Al$_{36}$Mg$_{52}$Zn$_{72}$ & 1  & 1   & 0    & $Im\bar{3}$ &-114.1&-109.0  \\  
Al$_{48}$Mg$_{64}$Zn$_{48}$ & 2/3 & 0   & 1/3 & $Immm$      &-92.8&-92.2  \\  
Al$_{66}$Mg$_{58}$Zn$_{36}$ & 1/4 & 1/2 & 0   & $R\bar{3}$  &-77.9&-76.1  \\  
\hline
\end{tabular}
\caption{The four stable low-temperature modifications of the 
Bergman phase $T$(AlMgZn): the occupation is given (as a fraction) 
of Zn on Al(2) site, Al on Mg(3) site and Al on Zn(1) site, 
followed by the space group.
Our site labels (Al/Mg/Zn) correspond to F/G/B in Ref.~\cite{mizutani}.
$\Delta E_{\rm VASP, pair}$ are the total energy (per atom, in eV) of each structure 
minus the reference (tie-plane) energy.}
\label{tab:almgzn}
\end{table}

%
%


\SAVE{Well in the future, a paper is planned by 
M. Mihalkovi\v{c}, M. de Boissieu, and others, on AlMgZn
(an expt MM participated in during 2007).  Not covered in this paper.}

\secc{Example 3: Al-Cu-Fe}
The structures of the best thermodynamically stable icosahedral quasicrystals, 
e.g. $i$-Al--Mn--Pd and $i$-Al--Cu--Fe, are poorly known 
despite excellent long--range order and 20 years of study.
This extends to $\alpha$-AlCuFe, the 
so-called 1/1 cubic approximant  to the quasicrystal
(it is related to $i$-AlCuFe 
in the same way that $T$(AlMgZn) [see above] is related to $i$-AlMgZn.)

Diffraction-based modeling, by itself, rarely resolves
all the important structural details; this is apparent even
in the $\alpha$-AlCuFe crystal, 
for which the ``solved'' structure~\cite{alpha-AlCuFeSi-expt}
contains mixed--occupancy Al--Cu and Al--Fe sites; that uncertainty, 
is a substantial obstacle to realistic modelling
of quasicrystal properties: the ab-initio energy of 
such a model is, we expect, unstable by $\sim$100 meV/atom
with respect to competing phases~\cite{AlCuFe-competing};
the idealized and improved model~\cite{alpha-AlCuFe-model} is,
we found, unstable by 51 meV/atom.
The root of the difficulty is that strong local correlations
have not been accounted for, due to the averaging inherent
in using Bragg peak intensities.

\begin{figure}
\includegraphics[width=3.0in]{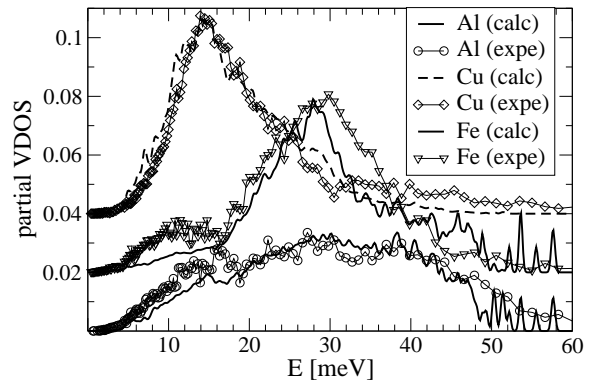}
\caption{Partial phonon densities of states for Al, Cu and Fe:``calc''
calculated (``calc'') using EOP potentials, experimental (``exp'')
data after Ref.~\cite{parshin}.}
\label{fig:alcufe-phonon}
\end{figure}

We set up a robust database with 15312
force and 120 energy datapoints (see Table~\ref{tab:database}). 
The fit converged to 0.2 eV/\AA~ r.m.s. for forces 
and 8  meV/atom for energies.
To refine the atomic structure of $\alpha(AlCuFe)$,
we took candidate sites from two diffraction-based models, 
either from  $\alpha$(AlCuFeSi)~\cite{alpha-AlCuFeSi-expt}, or
1/1-AlCuRu~\cite{alpha-AlCuRu-expt}, and performed
a lattice-gas Monte Carlo annealing using
EOP potentials. The best configurations were then annealed by 
MD (still using EOPP) at moderately high temperatures 
($T=$1000K) and finally quenched.
When their energies were recalculated ab-initio, the best example 
was still unstable, but only by the relatively small energy 35 meV/atom.
The lowest energy structure obtained as described above was then
used for phonon DOS calculation by diagonalizing the dynamical matrix.
The agreement between calculated and experimental DOS 
(Fig.~\ref{fig:alcufe-phonon}) is encouraging.

\SAVE{[MM-Jan-11  : In prev. paragraph,
do you mean (as I guessed) you took the union of atomic sites in the 
two structure models?
MMM: No, I didnt take the union, simply for historical reasons.
2 years ago I started working with AlCuFeSi-1/1, and obtained
at that time best solution +45 meV/atom above tie-plane. Recently in a new
round I studied AlCuRu, applied (about) the same procedure 
described above and obtained the best 1/1 so far. (As I mentioned
before, I also did few from-the-scratch trials, starting from random
set of points. The best I got has ab--initio recalculated energy at
+41meV/atom from tie-plane. The structure is not analyzed yet, I might
do so for my big paper.}

EOP potentials were also applied to predict the atomic structure of
the quasicrystal $i$-Al$_{67.5}$Cu$_{25}$Fe$_{12.5}$, by similar
lattice-gas approach using (a periodic approximant of) a ``quasilattice'' 
of candidate sites.~\cite{AlCuFe-pair}
The resulting structures were well-described by the formalism of a
cut through a six-dimensional hyper-structure, but the 
hyper-atom occupancies differ in detail from the well-known 
Katz-Gratias model~\cite{quiquandon}.
The ab-initio energy after relaxation was unstable 
(with respect to the competing phases~\cite{AlCuFe-competing})
by 45 meV/atom.
The computed vibrational DOS agrees with neutron results~\cite{parshin}.
\OMIT{(for each species)}

\SAVE{The lattice-gas sites are specified by an oversized 
Katz-Gratias acceptance domain, in fact.}

\secc{Conclusion.}
We have shown that empirical oscillating potentials
with the simple analytic form (\ref{eq:oscil6})
mimic the atomic interactions of many metallic
systems with sufficient accuracy to stand in for
ab-initio energies when those would be computationally
prohibitive, e.g. to resolve mixed
or partially occupied sites in diffraction refinements
of complex structures,
to measure thermodynamic expectations,
or to compute phonon spectra.

\SAVE{
It may seem surprising that the (comparatively) crude description
by pair potentials can sort correctly out energy differences down
to 1 meV/atom, when the {\it absolute} energies can be in
error (referenced to ab-initio) by $\sim 100$ times as much.
The key point is that the energy differences are small
because the structures being compared have 
{\it very similar} local order, and hence have the same systematic errors.
(The same might be said for LDA energies as referenced
to methods with better correlation energies.)}
\SAVE{Furthermore, it is known  that
the oscillating tails are crucial in deciding the competition
among quasicrystal structures and related crystals,
and this is likely true in other complex alloys.}

\SAVE{
Memo for some other paper (a technical detail):
A possible bottleneck with lattice-gas MC (of example 2, above): 
the refined sites are averaged and may not describe adequately all 
symmetry-breakings.  If the relaxed site
is significantly displaced from the average site,
it may have a spuriously low probability of occurring 
in the MC simulation.''}

The easily obtained EOP potentials may serve as 
a starting point for the (tedious) construction of 
(more elaborate) EAM versions of the potentials.
Furthermore, EOP potentials can accelerate ab-initio calculations,
to identify uncompetitive alternatives to be ignored
(as in our Al-Zn-Mg example)
or to carry out all but the final iterations of a relaxation.
\SAVE{MW notes:
The mGPT potentials certainly did aid initial relaxation 
for later ab-initio relaxation in the case of AlCoNi, so
MW thinks the EOP version of the same potentials is
also good for this purpose.  Has not been tried for Fe-B.}
EOP potentials appear to be fitted more robustly
than EAM potentials for 4-component systems;
we fitted B-C-Fe-Mo
which enables for the first time MD
simulations of technologically relevant metallic glasses 
(to probe the glass transition, two-level systems, etc.)
\SAVE{Ganesh says (5/2007) the swap rate acceptance for these 
quaternaries is still 0.5}.
It should be noted, however, that {\it any} pair potential 
will fail
in structures where the electron density has large variations 
in space (e.g. vacancies, edge dislocations, or surfaces).

\SAVE{MM-Jan'- Do we need to correct the referencing for B-C-Fe-Mo?
see MW and Ganesh emails of 1/4/08.
Ganesh email 1/4/08: the PRB cited doesn't discuss B-C-Fe-Mo;
MW: A J. Phys. Cond. Matt. paper (in preparation by MW) only mentions 
CBFeMo only very briefly, with a citation to the present paper! 
(See http://euler.phys.cmu.edu/widom/pubs/drafts/JPCM.pdf, 1/4/08).
MMM: lets stick to Mike's proposal - no explicit citation.
}

\SAVE{
Hennig had asked for a warning like the last sentence.
(But he told MM that such FIXED-VOLUME POTENTIALS are useful 
for studying SCREW DISLOCATIONS -- which don't have
excess free volume.  In fact, he
suggested MM try EOP potentials -- maybe to
probe screw dislocations -- in  Ti-based systems).
MM adds, the Ti case is an example where the usual
EAM totally fails: they had to explicitly include directionality.}

Rather than  emphasize the easy cases (alloys rich in simple metals, 
e.g. AlMgZn, Mg-X, but also Zn-Y),
we presented two difficult examples (B-Fe and Al-Cu-Fe),
which are borderline for pair potentials, 
due to covalent bonding of nearest neighbor transition metals.
In the many cases where bond directionality is even more important, 
EOP potentials cannot be used. 

\SAVE{MM notes: (1) It's surprising that a pair-potential works at all
in view of the non-negligible magnetic interactions!
(2) We selected the Al--Cu--Fe ternary case here being well aware that
the pair interactions cannot offer accurate description of the
interactions at such substantial fractional content of Cu atoms, with
important Cu--Cu and Cu--Fe interactions. However, we would hope that
they would be accurate enough to provide useful and cheap constraints
to guide subsequent ab--initio calculations.}

\SAVE{MM notes (from Ref.~\onlinecite{MMreview}: Ir and Rh are known to 
be not fittable by simple EAM, because of [bond] directionality 
(there is an explicit argument how the directionality is evidenced 
in EAM setup, something with curvature of the embedding function). 
Out of curiosity, MM tried to fit Mg-Ir system with EOPP;
indeed, the fit is very bad, in contrast to Mg-Zn, Mg-Pd, Mg-Ag...}

\SAVE{Our Mg-Mg potential comes out virtually identical with the
``GPT'' potential~\cite{moriarty-GPT}(b);
our Al-Al fit agrees with GPT in the oscillating part, but the
shallow repulsive shoulder is $\sim$$20$ meV lower than 
in the GPT case.}

Our method had mixed success for pure elements:
Al-Al potentials show an excellent fit to the ab-initio data, with
Mg-Mg even better.  
But with pure Zn, 
\OMIT{even with better functional forms than  (\ref{eq:oscil6}),}
pair potentials never reproduced the ab-initio phonon vibrational DOS;
and the EOPP approach fails for pure Ga.
\SAVE{MM-12'' (still)   
It would be better to somehow be quantitative on the 
agreement between EO and ab-initio; can't we say something like
dE(EO-vasp)/dE(GPT-vasp) $\sim$ 0.5?
Or were you saying GPT is the gold standard, so the closer
we are to GPT the better? I ask this, because I noticed that
Mg is closer to vasp than Al is.
MMMM: My recollection (not digging into old data now) is that GPT
have usually starting error in forces ~ 0.3-0.4 eV/A. Fitted EO
gets that down to 0.1-0.2 eV/A. However, my feeling is this is
underestimate of goodness of GPT: usually our datasets contain
at least few awkward samples, and these add a lot to GPT error,
why they get fixed by fitting procedure for EO. 
MM: just turning this into SAVE.
}

EOP potentials show promise for many other alloy systems.
They allowed extraction of
long-wavelength phonons from molecular dynamics simulations~\cite{bili}
in liquid Bi--Li system.
Outstandingly accurate EOP were fitted for the system
Mg--T (T=Pd or Ag) in the Mg-rich limit;
for T=Ag, the Ag-Ag potential exhibits strong 
oscillations to large $r$ that could only be fitted after the cutoff 
was extended to to 14\AA~\cite{kreiner}.
These systems include various complex phases based on icosahedral
``Mackay'' clusters  (Mg$_{42}$X$_{12}$).
EOP potentials have predicted -- correctly -- which Mg sites 
get substituted by Pd as the composition is varied~\cite{kreiner}
in the phases $\gamma$(Mg$_{0.8}$Pd$_{0.2}$)
and $\delta$(Mg$_{0.8}$Pd$_{0.2}$) [both ``1/1 approximants'', 
similar in structure to $\alpha$-AlMnSi].
\SAVE{
There is a finite range of stability in these Al-Mg structures,
which are MI based, with Mg playing the role of Al.
Both $\gamma$ and $\delta$ are 1/1 approximant, and have virtually same 
composition;
$\gamma$ is like alpha-AlMnSi ($Pm\bar{3}$ spacegroup), $\delta$
is disordered with $Im\bar{3}$ spacegroup. Actually, 
both these phases are high-temperature, so it is a tricky and subtle 
issue to explain stability of $\gamma$ which does not have reported disorder.}


\SAVE{The D-W was accurately described, 
including the ``static'' component due to disorder.
What this means (MM 1/08):
Imagine you have fractional/mixed occupancy of an orbit. Then you can
compute rms displacement corresponding to the spread of the points around
the symmetry-adjusted mean (like in tiling-deco. models!) - this is the
``static'' DW factor. After one computes the proper phonon DW and adds the
static DW, that should be the expe. measured DW factor. Accuracy of
these calculations is really remarkable, the results agree quantitatively.}

\SAVE{Another interesting system:
``There are inviting prospects to employ EOP potentials 
(i) in the Zn-Y (Zn-rich) system, for the correct description of Zn-rich 
phases. There are nine(!) experimentally observed stable
or metastable phases, for Y fractional content up to $x_Y$=0.25.
Also (ii) for the case Mg-Zn-Y: there is an excellent force AND energy fit.''
[rms forces are
0.10 eV/\AA~ and rms $\Delta E\sim$3 meV/atom for 114 $\Delta E$ datapoints.]
However, applications haven't yet been carried out.}

\acknowledgments
We thank R. G. Hennig and F. G\"ahler for comments.
We supported by DOE Grant DE-FG02-89ER-45405
(MM, CLH),
the DARPA Structural Amorphous Metals Program under 
ONR Grant N00014-06-1-0492 (MW, MM, PG), and
Slovak funding VEGA 2/0157/08 and APVV-0413-06 (MM).


\begin{thebibliography}{99}
\bibitem{fn-densityrange}
For a successful fit, the database 
must be limited to a relatively
narrow range of (electron) densities, so the Fermi wavevector 
[hence $k_*$ in Eq.~(\ref{eq:oscil6})] has a consistent value.
\SAVE{Or course, the resulting validity will be
limited to the same narrow range.}
\bibitem{FN-need-energies}
Fits from forces alone were insufficiently constrained;
including energy differences, even with low weight,
improved the fit.
\end{thebibliography}
\end{document}